\documentclass[twocolumn]{revtex4-1}
\usepackage[utf8]{inputenc}
\usepackage{graphicx}
\usepackage{amsmath}
\usepackage{amsfonts}
\usepackage{amssymb}
\usepackage[left=2cm,right=2cm,top=2cm,bottom=2cm]{geometry}

\begin{document}
\title{Hubbert's theory and photovoltaics: the nonsense race of breaking energy-conversion records}
\date{\today}
\author{Marcos Paulo Belançon}
\email{marcosbelancon@utfpr.edu.br}
\affiliation{Programa de Pós-graduação em Processos Químicos e Bioquímicos\\Universidade Tecnológica Federal do Paraná\\CEP 85503-390, Pato Branco, Paraná, Brasil}

\begin{abstract}
The wide public sees solar energy as the future of mankind, and media channels quite oftenly states that our challenge is to improve efficiencies and reduce cost. However, one may point some unconvenient truth's about the physical limits we are facing, that are barely discussed by the public or even by scientists and institutions that are strongly biased towards a picture of a sustainable oil free energy in the future. In this work we discuss some of those physical limits of photovoltaics based on the principle of the Hubbert's theory for the oil peak, evidencing that much of the research is focused on photovoltaic efficiencies and this parameter is widely overestimated: better efficiencies oftenly are the result of complex technologies that are expensive and not scalable. In this context, if fossil fuels proved not replaceable, it is very likely that our socioeconomic ideas based in the past will not withstand an energy transition to high cost and low quality sources. 
\end{abstract}

\maketitle

\section{Improving photovoltaics}

\begin{figure}[h]
\center
\includegraphics[scale=0.3]{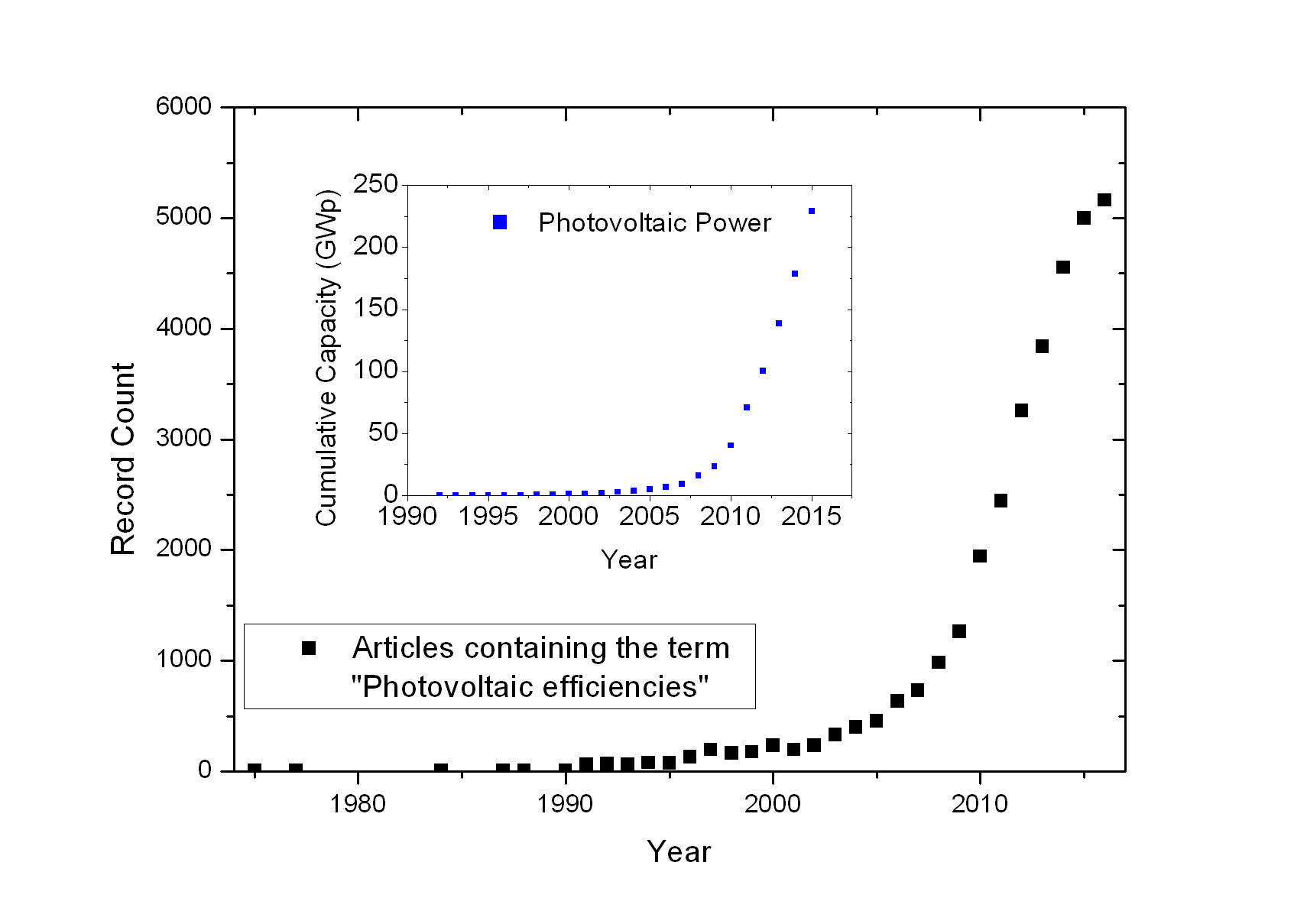}
\caption{Number of articles in the Web Of Science database containing the term ``photovoltaic efficiencies''.}
\label{WOS}
\end{figure}

A quick search in the \textit{Web Of Science} for ``photovoltaic efficiencies'' provides us an insight of how much research is being carried on this subject, or is at least related to it. This data is shown in figure \ref{WOS}, where in the inset we can see the cumulative photovoltaic (PV) power in the world.

As one can see, the same exponential behavior is observed in both curves. However, despite growth records of PV's in the world, forecasts from the \textit{Intergovernmental Panel on Climate Change (IPCC)}, the \textit{Energy Information Administration (EIA)} or from companies such as the \textit{British Petroleum (BP)} still give a picture of a future based mainly on oil and natural gas. Some will say that the potential of solar energy is being underestimated\cite{Creutzig2017}.

Many different dreams about our sustainable energy future are being discussed\cite{Chu2012,Haegel2017}. Thousands of researchers are trying to improve solar energy to electricity conversion today; about 14 articles are being published everyday covering from development of new PV's\cite{Lee2016,Albrecht2017} until spectral converters\cite{Huang2013a} and solar thermal devices\cite{Kannan2016,NathanSLewis2016}. A handful of different PV technologies are already very competitive in terms of efficiency\cite{Rand2007,Summary2016}, however, as it is indicated in figure \ref{WOS} our community is worried in further improvements.

\section{The role of energy-conversion efficiencies}

The unconvenient truth is that PV efficiencies may not help our energy transition. Looking for example to the \textit{National Renewable Energy Laboratory (NREL)} efficiency chart for PV's, one may see that the Silicon technology widely dominant in the market today is almost 30 years old. By this way one may interprete that, except by improvements in production techniques, Silicon PV's are still the same. When scientists have developed a new PV technology in the last few decades, even though some energy conversion records were broken, it has produced no net effect in our hability to produce solar energy. Some technologies, such as CdTe PV`s are competitive today, however the dependende of scarse minerals makes impossible to scale up the production as much as we need.

Even more interesting is to have a look in the NREL PV System Cost Benchmark\cite{Fu2017} where they modeled the impact of module energy-conversion efficiency ($\eta$) on total system costs. While authors oftenly interpret that improve $\eta$ is ``probably the key''\cite{Green2016} to reduce costs, the NREL model, even considering the optimistic assumption of reach $\eta=60\%$ by the same price of today's modules, does not project a strong reduction in the total cost. As the data in table \ref{tabelatempos} shows, higher efficiencies may reduce costs by only $\sim 20\%$.

\begin{table}[h]
\centering
\begin{tabular}{|c|c|c|c|}
\hline
Type&\bf $\eta$ & Cost (US\$/Wp)& Cost if $\eta=60\%$\\
\hline \hline
Residential & 16.2\%&2.90 &2.18\\ 
\hline 
Commercial & 17.5\%&1.84 &1.31\\ 
\hline 
Utility-Scale & 17.5\%&1.09 &0.78\\ 
\hline 
\end{tabular} 
\caption{Costs and average efficiencies of different PV's projects \footnote{data from \cite{Fu2017}, pg 45.}}
\label{tabelatempos}
\end{table}

This data should not be miss interpreted: higher $\eta$ itself is desirable, however, if we aim an approach towards global scale applications of PV's, one may not neglect the scale. 

Two centuries ago, the economist Jean-Baptiste Say\cite{say1803} assumed that the scale of our consumption was so small that natural resources would last forever. This assumption, that natural resources are not produced or consumed is the basis of his work. It is remarkable that this aproximation made by Say has worked by so much time, however, everyone should consider that since his time we multiplied the population several times and the energy consumption per capita even more. 

Many natural resources are now being consumed, such as oil and gas; many others are being diluted, such as Silver, Copper and every single element mined on earth. Mining or recycling less concentrated ores consumes more energy, what is a consequence of the laws of thermodynamics. Modern economies worldwide are not considering the implications that today's reality is imposing to us, and it does not matter if we peak up a liberalist or communist  one; the economic theories behind will always consider that the market will find another resource once the first one is exhausted.

\section{Why scale matters?}

\begin{figure}[h]
\center
\includegraphics[scale=0.3]{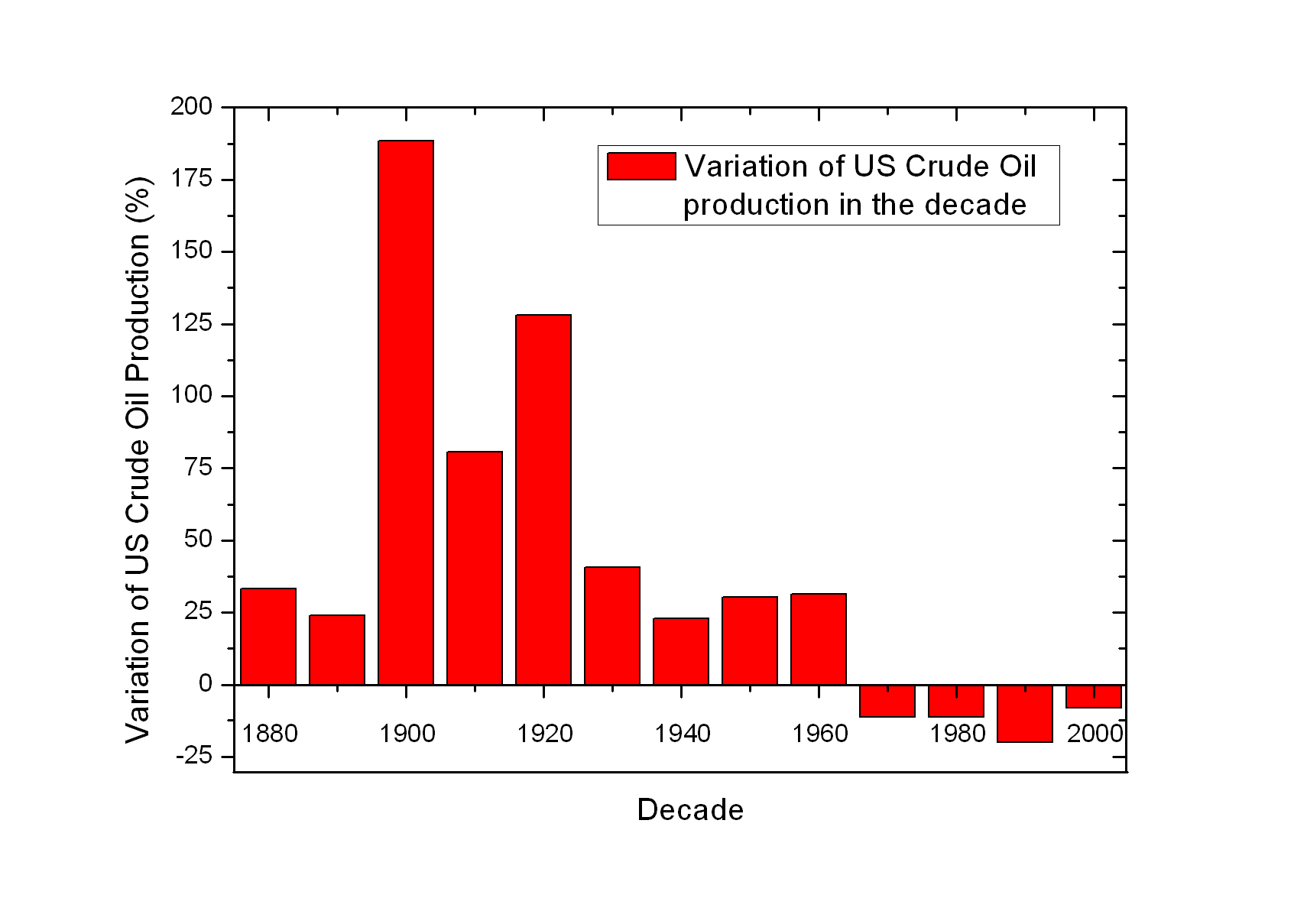}
\caption{Percentual growth of US oil production per decade.}
\label{crude}
\end{figure}

Even though PV's cumulative capacity growth may be interpreted as astoshing or promising, one should take care to make a fair comparison between the facts and our expectations. To give some perspective, in figure \ref{crude} we have the variation of US crude oil production in each decade since the 1880's.

From the begining of the oil exploration until the peak of conventional oil production in the 1970's the producion has grown pretty fast. Between 1900 and 1929 the production was multiplied by a factor of 15. And then, after the peak in the 1970's, oil imports contributed to install a chronic debt in the trade balance of the country with the world. Even though the US had already known non-conventional oil and gas sources, the market has choose to buy the cheaper conventional alternative in Venezuela and Midde-east.

It is important to pay attention to some facts in this history. When the cumulative capacity is very low it is common to have growth rates of two or even three digits. In 1900 the US oil production has grown at rates similar to those predicted for the photovoltaics in the coming decade. The other lesson from the US oil peak is that consequences of such event may be huge and difficult to predict, economically and politically.

\begin{figure}[h]
\center
\includegraphics[scale=0.3]{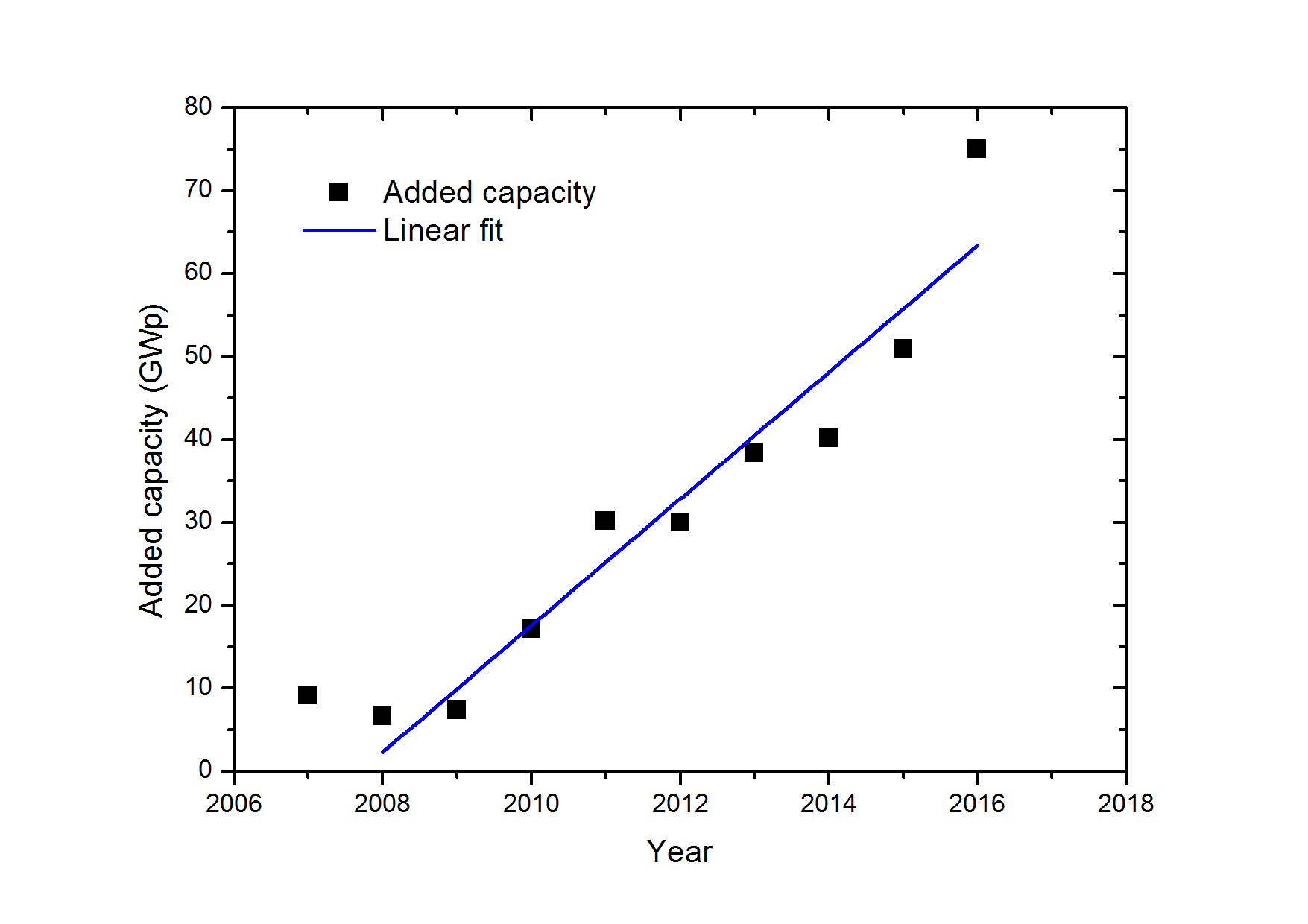}
\caption{PV's world production capacity.}
\label{pc}
\end{figure}

Hubbert predicted in 1956\cite{Hubbert1956} the decline of oil production in US and 15 years after the history proved him to be right. Hubbert's theory can not be applied to solar energy, however, it may be more or less applied to all the raw material we need to build the PV's. By this way, we should take a look in the real scale of supply, demand and raw material consumption for solar energy. Figure \ref{pc} shows a plot of the added capacity (AC) in the last few years, from which we may found that it has grown more or less linearly. 

The challenge that PV's are facing is not only in reducing cost or improving efficiencies; neither is increase production capacity. The challenge is if AC can increase faster than linearly, or at least what slope is sustainable from the perspective of resources and cost. In the last few year the prices were pressured down mainly by  overproduction in China, what put many companies in the west in a difficult situation.

On the perspective of available resources, the \textit{World Silver Survey 2017}, from \textit{The Silver Institute}, shows that in 2011 and in 2016 the PV's industry has consumed the same amouth of Silver, of about 75 million ounces, while the AC speed up by $\sim3$. This achievement is remarkable, however, as it shown in \textit{International Technology Roadmap For Photovoltaics 2017}, it is not so clear that industry will continue to reduce Silver consumption per cell so fast as in the past. There are not much room for improvements when the technology is already mature, and Silver production is a concern\cite{Sverdrup2014,Grandell2014}.

The world PV's installed capacity is around 300 GWp today, which corresponds to an average of 60 GW of electricity. This is in the range of the the energy equivalent to the US oil production in the time of the first world war, which was less than 1 million barrels a day. These data discussed here supports the idea that even running far from the scale of mankind's needs, PV's are already facing finite resources constraints\cite{Tao2011}.

After China dominated the PV's market, many companies in US, Europe and Japan lost value or even declared insolvency. To keep AC increasing while maintaining or reducing the prices the PV's industry needs to keep sharing more or less the same 75 million ounces of Silver they consumed last year. In other words, it is necessary to reduce the silver content\cite{Grandell2014}. From the materials perspective Copper may be used to replace Silver, however it does not have the same conductivity and chemical resistance. The assumption that we will use less materials, with lower quality but engineered in such way that will result in more efficient devices is too much optimistic and not scientific.

Back to the scale needed, even a conservative projection\cite{FraunhoferISE2015} considers an AC of PV's per year three times higher than today. While Silicon technology needs Silver, others are based on elements even more difficult to obtain, such as Cadmium, Tellurium, Indium, Gallium, Selenium or some combination of them. A few of those elements are byproduct mineral\cite{Bleiwas2010,Graedel2011,Grandell2016} and by this way the production of such CdTe, CIS or CIGS PV's is very unlikely to reach the Silicon PV's production scale\cite{Tao2011}. If our best shot to produce solar electricity will be with the 30 years old single junction Silicon technology, we may reduce Silver consumption per cell further, what will enable AC to be increased by the ``conservative'' (or realistic) factor of three in the next decades.

\section{Production capacity in the future}

\begin{figure}[h]
\center
\includegraphics[scale=0.25]{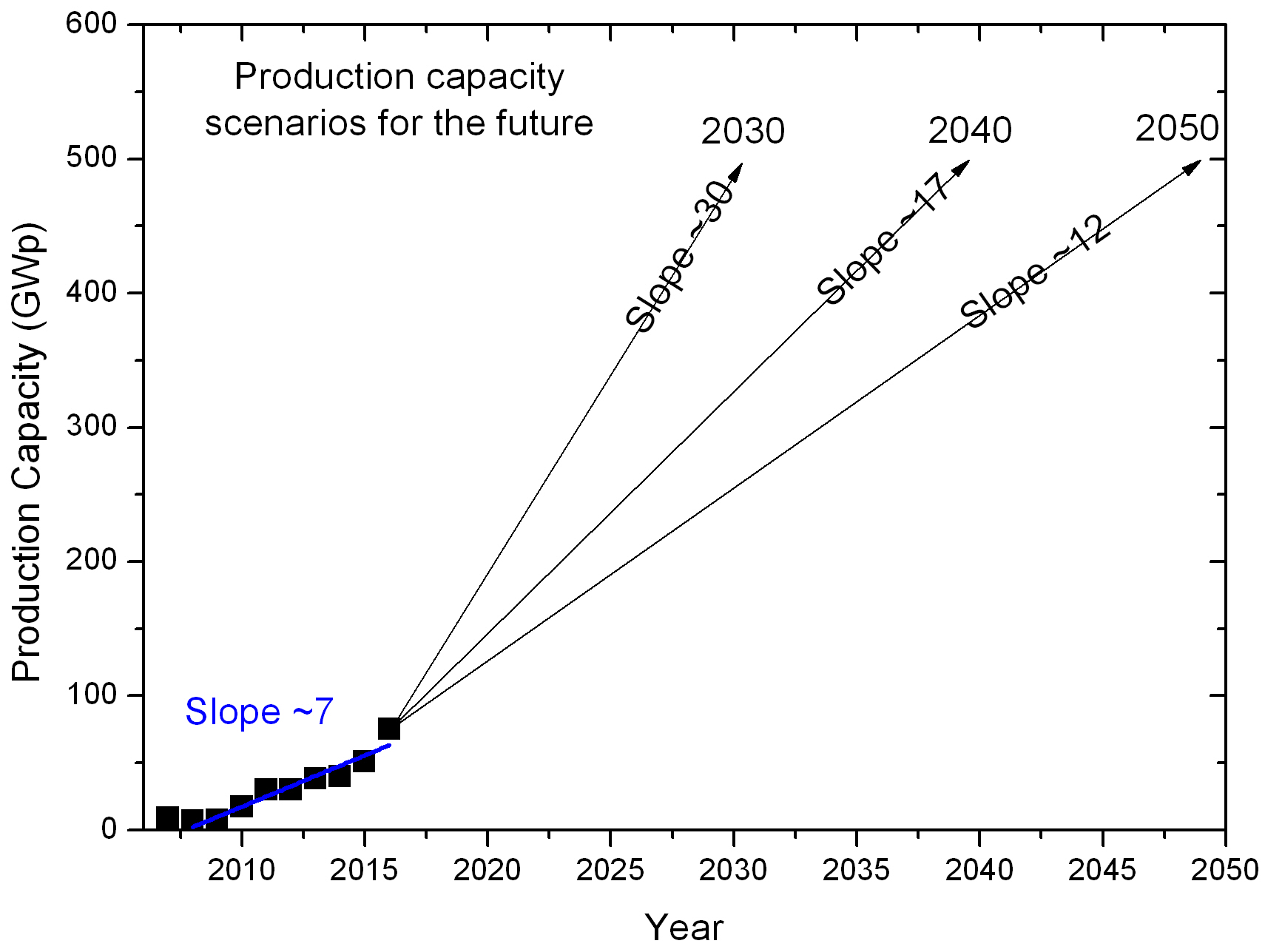}
\caption{PV's world production capacity and slopes needed to reach 500 GWp in the next decades.}
\label{projection}
\end{figure}

To clarify this picture of how far we are from the ``Hubbert's limits" the figure \ref{projection} show some very simple scenarios for PC of PV's. At left in the figure we can see the same data from figure \ref{pc}, for the AC, as well the linear fit represented by the blue line. The slope for this curve is about 7 GWp/year. This is how much we are speeding up the addition of PV`s in the world in the last 10 years, on average. If the slope surpasses the growing demand of PV`s, we should have and excess suply that in will in turn make some companies to crash: a common market driven adjustment. 

The three other lines in the figure \ref{projection} project what should be the slope if we aim to reach $PC=500$ GWp in 2030, 2040 or 2050. This 500 GWp of PC should be enough to reach in a few decades about 12.5 TWp of installed capacity; which means something like 2.5 TW on average. This value is still a fraction of how much energy we consume worldwide but it is useful to provide a picture of the point discussed here.

If PC keeps increasing at the present rate, we are not going to reach the 500 GWp before 2060. Even to reach this level in 2050 the slope needs to be increased to about 12 GWp/year. This means for example that, to avoid Silver constraints the industry should reduce its consumption per Wp in about 80\% in 33 years. Even though this kind of reduction was achieved in the last ten years, it is not clear that such reduction can be achieved again without undesirable performance effects. And this is only one example.

The point here is that the question that should receive more attention from the community is if are all the pieces in the production chain of photovoltaics ``free" to grow? Every time one constraint appear, the prices may not decrease, what can reduce demand and investment, compromising the PC growth. Our biggest challenge is than the scale of our needs and a quite short deadline.

\section{Scaling up production instead of improving efficiencies}

In a recent report from the HSBC Global Research\cite{Fustier2017}, Fustier et al made estimates that it is very likely that today's oil field will decline in production by more than 40 million barrels a day until 2040. To replace it, the non-conventional fossil fuels are already been extracted, mainly in Shale plays in US\cite{Hoegh-guldberg2007,Hughes2013} and oil sands in Canada. 

Since 2008 the shale rush made the US Oil production increase from $\sim5$ to $\sim9$ million barrels a day, deviating from the Hubbert prediction made 70 years ago. 4 million barrels more per day is energy equivalent to much more than what PV's of the world can deliver today. However, this non-conventional oil source has its own problems: the decline production of individual Shale wells may be as fast as 50\% in the first year. So, the Hubbert's peak for this non-conventional source is very likely to come faster than for conventional sources. By this way, the oil decline is very likely do impact the cost of our energy at the same time it may introduce a kind of deadline to perform a transition to renewables.

I wish PV's could be the alternative to replace conventional oil and gas. However, considering the prediction made by Fustier, to accomplish that we should multiply our installed PV's by a factor of about 40 until 2040, which should require a PC of 500 GWp on average from now on; which is clearly not feasible. Anyway, if such oil decline is going to happen, to replace it scientists working in the field should change their research focus.

Across the globe we are working in ``photovoltaic efficiencies'' as mentioned earlier, and it is very unlikely that such contributions will ever have some effect in our everyday lives. We should than focus on expanding the limits of the cumulative capacity, by increasing life span of PV's, reducing and eventually replacing scarse materials such as Silver and developing techniques that could make recycling PV's easier.

All of that could make possible to achieve a higher slope in PC growth; increase photovoltaic efficiences does not help pretty much towards this direction because we are already near the limits of single junction solar cell efficiencies. And we may add that it could be helpful for scientists and the public in general to think more broadly about the energy transition. If replace fossil fuels prove not to be possible in the coming decades, we may need to look to other options.

\section{Other option?}

The mankind's appetite for energy is huge and our institutions are also addicted to it. The IPCC has more specialists in economics than in ecology, and the prevalence of men and europeans is clear\cite{VandenBergh2017a}. When scientists try to follow the biased view of this kind of institution to pursue grants and publications, science as a whole is in serious danger. The IPCC does not put all option in the table (and in the reports), because only perpetual growth of gross domestic product (GDP) are allowed to be present in the debate. A possible limit to the ideology of unlimited growth\cite{Kallis2011,VandenBergh2017a} is not being considered, even though it could solve many aspects of our situation.

Once we are not near of solving our sustainability problem\cite{Raftery2017}, all the options should be present in the scientific debate promoted by institution such as the IPCC. That should include many aspects of the feasibility in the long term of our civilization as it is today. There are socioeconomic options that aim reduce consumption of resources and increase prosperity\cite{timjackson}, however, the institutions did not consider it seriously.

\section{Conclusion}

If we choose to replace conventional by non-conventional fossil fuels, the overall cost of energy will rise. On the other hand, if we choose renewables, such as PV`s, to reach the scale of our needs it is very likely that the overall cost of energy will rise, also.

While scientists are blindly wasting time following what may be proved as useless goals, they are also wasting intellectual resources that could be used to fix our civilization instead of, at maximum, create a new PV company to produce a few more GW's that may solve a tiny part of the problem. The tragedy of pursuing continuous growth of the energy supply is that the manutention costs of our civilization are growing together\cite{Tainter2012}; and after the Hubbert's peak of the key resource we will face deplection while our manutention cost keeps rising. There are no high tech solution that will avoid this dilemma. 

In the 1980's France has built\cite{Dussud2016} an impressive fleet of nuclear reactors, and even though Uranium production did not peaked, it did not avoid the problem that now the country needs to replace the nuclear reactors to keep the supply of energy. There are other hundreds of nuclear reactors facing the same critical point around the world.

So, the big question is: can we replace fossil fuels by PV`s or other renewables without replace our socioeconomic status quo? Some scientists will answer that yes, of course. But these answers are about faith and not supported by scientific evidence.

\bibliography{/home/mbelancon/Documentos/library}

\begin{thebibliography}{30}%
\makeatletter
\providecommand \@ifxundefined [1]{%
 \@ifx{#1\undefined}
}%
\providecommand \@ifnum [1]{%
 \ifnum #1\expandafter \@firstoftwo
 \else \expandafter \@secondoftwo
 \fi
}%
\providecommand \@ifx [1]{%
 \ifx #1\expandafter \@firstoftwo
 \else \expandafter \@secondoftwo
 \fi
}%
\providecommand \natexlab [1]{#1}%
\providecommand \enquote  [1]{``#1''}%
\providecommand \bibnamefont  [1]{#1}%
\providecommand \bibfnamefont [1]{#1}%
\providecommand \citenamefont [1]{#1}%
\providecommand \href@noop [0]{\@secondoftwo}%
\providecommand \href [0]{\begingroup \@sanitize@url \@href}%
\providecommand \@href[1]{\@@startlink{#1}\@@href}%
\providecommand \@@href[1]{\endgroup#1\@@endlink}%
\providecommand \@sanitize@url [0]{\catcode `\\12\catcode `\$12\catcode
  `\&12\catcode `\#12\catcode `\^12\catcode `\_12\catcode `\%12\relax}%
\providecommand \@@startlink[1]{}%
\providecommand \@@endlink[0]{}%
\providecommand \url  [0]{\begingroup\@sanitize@url \@url }%
\providecommand \@url [1]{\endgroup\@href {#1}{\urlprefix }}%
\providecommand \urlprefix  [0]{URL }%
\providecommand \Eprint [0]{\href }%
\providecommand \doibase [0]{http://dx.doi.org/}%
\providecommand \selectlanguage [0]{\@gobble}%
\providecommand \bibinfo  [0]{\@secondoftwo}%
\providecommand \bibfield  [0]{\@secondoftwo}%
\providecommand \translation [1]{[#1]}%
\providecommand \BibitemOpen [0]{}%
\providecommand \bibitemStop [0]{}%
\providecommand \bibitemNoStop [0]{.\EOS\space}%
\providecommand \EOS [0]{\spacefactor3000\relax}%
\providecommand \BibitemShut  [1]{\csname bibitem#1\endcsname}%
\let\auto@bib@innerbib\@empty
\bibitem [{\citenamefont {Creutzig}\ \emph {et~al.}(2017)\citenamefont
  {Creutzig}, \citenamefont {Agoston}, \citenamefont {Goldschmidt},
  \citenamefont {Luderer}, \citenamefont {Nemet},\ and\ \citenamefont
  {Pietzcker}}]{Creutzig2017}%
  \BibitemOpen
  \bibfield  {author} {\bibinfo {author} {\bibfnamefont {F.}~\bibnamefont
  {Creutzig}}, \bibinfo {author} {\bibfnamefont {P.}~\bibnamefont {Agoston}},
  \bibinfo {author} {\bibfnamefont {J.~C.}\ \bibnamefont {Goldschmidt}},
  \bibinfo {author} {\bibfnamefont {G.}~\bibnamefont {Luderer}}, \bibinfo
  {author} {\bibfnamefont {G.}~\bibnamefont {Nemet}}, \ and\ \bibinfo {author}
  {\bibfnamefont {R.~C.}\ \bibnamefont {Pietzcker}},\ }\href {\doibase
  10.1038/nenergy.2017.140} {\bibfield  {journal} {\bibinfo  {journal} {Nature
  Energy}\ }\textbf {\bibinfo {volume} {2}},\ \bibinfo {pages} {17140}
  (\bibinfo {year} {2017})}\BibitemShut {NoStop}%
\bibitem [{\citenamefont {Chu}\ and\ \citenamefont {Majumdar}(2012)}]{Chu2012}%
  \BibitemOpen
  \bibfield  {author} {\bibinfo {author} {\bibfnamefont {S.}~\bibnamefont
  {Chu}}\ and\ \bibinfo {author} {\bibfnamefont {A.}~\bibnamefont {Majumdar}},\
  }\href {\doibase 10.1038/nature11475} {\bibfield  {journal} {\bibinfo
  {journal} {Nature}\ }\textbf {\bibinfo {volume} {488}},\ \bibinfo {pages}
  {294} (\bibinfo {year} {2012})}\BibitemShut {NoStop}%
\bibitem [{\citenamefont {Haegel}\ \emph {et~al.}(2017)\citenamefont {Haegel},
  \citenamefont {Margolis}, \citenamefont {Buonassisi}, \citenamefont
  {Feldman}, \citenamefont {Froitzheim}, \citenamefont {Garabedian},
  \citenamefont {Green}, \citenamefont {Glunz}, \citenamefont {Henning},
  \citenamefont {Holder}, \citenamefont {Kaizuka}, \citenamefont {Kroposki},
  \citenamefont {Matsubara}, \citenamefont {Niki}, \citenamefont {Sakurai},
  \citenamefont {Schindler}, \citenamefont {Tumas}, \citenamefont {Weber},
  \citenamefont {Wilson}, \citenamefont {Woodhouse},\ and\ \citenamefont
  {Kurtz}}]{Haegel2017}%
  \BibitemOpen
  \bibfield  {author} {\bibinfo {author} {\bibfnamefont {N.~M.}\ \bibnamefont
  {Haegel}}, \bibinfo {author} {\bibfnamefont {R.}~\bibnamefont {Margolis}},
  \bibinfo {author} {\bibfnamefont {T.}~\bibnamefont {Buonassisi}}, \bibinfo
  {author} {\bibfnamefont {D.}~\bibnamefont {Feldman}}, \bibinfo {author}
  {\bibfnamefont {A.}~\bibnamefont {Froitzheim}}, \bibinfo {author}
  {\bibfnamefont {R.}~\bibnamefont {Garabedian}}, \bibinfo {author}
  {\bibfnamefont {M.}~\bibnamefont {Green}}, \bibinfo {author} {\bibfnamefont
  {S.}~\bibnamefont {Glunz}}, \bibinfo {author} {\bibfnamefont {H.-m.}\
  \bibnamefont {Henning}}, \bibinfo {author} {\bibfnamefont {B.}~\bibnamefont
  {Holder}}, \bibinfo {author} {\bibfnamefont {I.}~\bibnamefont {Kaizuka}},
  \bibinfo {author} {\bibfnamefont {B.}~\bibnamefont {Kroposki}}, \bibinfo
  {author} {\bibfnamefont {K.}~\bibnamefont {Matsubara}}, \bibinfo {author}
  {\bibfnamefont {S.}~\bibnamefont {Niki}}, \bibinfo {author} {\bibfnamefont
  {K.}~\bibnamefont {Sakurai}}, \bibinfo {author} {\bibfnamefont {R.~A.}\
  \bibnamefont {Schindler}}, \bibinfo {author} {\bibfnamefont {W.}~\bibnamefont
  {Tumas}}, \bibinfo {author} {\bibfnamefont {E.~R.}\ \bibnamefont {Weber}},
  \bibinfo {author} {\bibfnamefont {G.}~\bibnamefont {Wilson}}, \bibinfo
  {author} {\bibfnamefont {M.}~\bibnamefont {Woodhouse}}, \ and\ \bibinfo
  {author} {\bibfnamefont {S.}~\bibnamefont {Kurtz}},\ }\href {\doibase
  10.1126/science.aal1288} {\bibfield  {journal} {\bibinfo  {journal}
  {Science}\ }\textbf {\bibinfo {volume} {356}},\ \bibinfo {pages} {141}
  (\bibinfo {year} {2017})}\BibitemShut {NoStop}%
\bibitem [{\citenamefont {Lee}\ and\ \citenamefont {Ebong}(2016)}]{Lee2016}%
  \BibitemOpen
  \bibfield  {author} {\bibinfo {author} {\bibfnamefont {T.~D.}\ \bibnamefont
  {Lee}}\ and\ \bibinfo {author} {\bibfnamefont {A.~U.}\ \bibnamefont
  {Ebong}},\ }\href {\doibase 10.1016/j.rser.2016.12.028} {\bibfield  {journal}
  {\bibinfo  {journal} {Renewable and Sustainable Energy Reviews}\ }\textbf
  {\bibinfo {volume} {70}},\ \bibinfo {pages} {1286} (\bibinfo {year}
  {2016})}\BibitemShut {NoStop}%
\bibitem [{\citenamefont {Albrecht}\ and\ \citenamefont
  {Rech}(2017)}]{Albrecht2017}%
  \BibitemOpen
  \bibfield  {author} {\bibinfo {author} {\bibfnamefont {S.}~\bibnamefont
  {Albrecht}}\ and\ \bibinfo {author} {\bibfnamefont {B.}~\bibnamefont
  {Rech}},\ }\href {\doibase 10.1038/nenergy.2016.196} {\bibfield  {journal}
  {\bibinfo  {journal} {Nature Energy}\ }\textbf {\bibinfo {volume} {2}},\
  \bibinfo {pages} {16196} (\bibinfo {year} {2017})}\BibitemShut {NoStop}%
\bibitem [{\citenamefont {Huang}\ \emph {et~al.}(2013)\citenamefont {Huang},
  \citenamefont {Han}, \citenamefont {Huang},\ and\ \citenamefont
  {Liu}}]{Huang2013a}%
  \BibitemOpen
  \bibfield  {author} {\bibinfo {author} {\bibfnamefont {X.}~\bibnamefont
  {Huang}}, \bibinfo {author} {\bibfnamefont {S.}~\bibnamefont {Han}}, \bibinfo
  {author} {\bibfnamefont {W.}~\bibnamefont {Huang}}, \ and\ \bibinfo {author}
  {\bibfnamefont {X.}~\bibnamefont {Liu}},\ }\href {\doibase
  10.1039/C2CS35288E} {\bibfield  {journal} {\bibinfo  {journal} {Chem. Soc.
  Rev.}\ }\textbf {\bibinfo {volume} {42}},\ \bibinfo {pages} {173} (\bibinfo
  {year} {2013})}\BibitemShut {NoStop}%
\bibitem [{\citenamefont {Kannan}\ and\ \citenamefont
  {Vakeesan}(2016)}]{Kannan2016}%
  \BibitemOpen
  \bibfield  {author} {\bibinfo {author} {\bibfnamefont {N.}~\bibnamefont
  {Kannan}}\ and\ \bibinfo {author} {\bibfnamefont {D.}~\bibnamefont
  {Vakeesan}},\ }\href {\doibase 10.1016/j.rser.2016.05.022} {\bibfield
  {journal} {\bibinfo  {journal} {Renewable and Sustainable Energy Reviews}\
  }\textbf {\bibinfo {volume} {62}},\ \bibinfo {pages} {1092} (\bibinfo {year}
  {2016})}\BibitemShut {NoStop}%
\bibitem [{\citenamefont {{Nathan S. Lewis}}(2016)}]{NathanSLewis2016}%
  \BibitemOpen
  \bibfield  {author} {\bibinfo {author} {\bibnamefont {{Nathan S. Lewis}}},\
  }\href {\doibase 10.1126/science.aad1920.22} {\bibfield  {journal} {\bibinfo
  {journal} {Science Research}\ }\textbf {\bibinfo {volume} {351}},\ \bibinfo
  {pages} {aad1920} (\bibinfo {year} {2016})}\BibitemShut {NoStop}%
\bibitem [{\citenamefont {Rand}\ \emph {et~al.}(2007)\citenamefont {Rand},
  \citenamefont {Genoe}, \citenamefont {Heremans},\ and\ \citenamefont
  {Poortmans}}]{Rand2007}%
  \BibitemOpen
  \bibfield  {author} {\bibinfo {author} {\bibfnamefont {B.~P.}\ \bibnamefont
  {Rand}}, \bibinfo {author} {\bibfnamefont {J.}~\bibnamefont {Genoe}},
  \bibinfo {author} {\bibfnamefont {P.}~\bibnamefont {Heremans}}, \ and\
  \bibinfo {author} {\bibfnamefont {J.}~\bibnamefont {Poortmans}},\ }\href
  {\doibase 10.1002/pip} {\bibfield  {journal} {\bibinfo  {journal} {Prog.
  Photovolt: Res. Appl.}\ }\textbf {\bibinfo {volume} {15}},\ \bibinfo {pages}
  {659} (\bibinfo {year} {2007})},\ \Eprint {http://arxiv.org/abs/1303.4604}
  {arXiv:1303.4604} \BibitemShut {NoStop}%
\bibitem [{\citenamefont {Polman}\ \emph {et~al.}(2016)\citenamefont {Polman},
  \citenamefont {Knight}, \citenamefont {Garnett}, \citenamefont {Ehrler},\
  and\ \citenamefont {Sinke}}]{Summary2016}%
  \BibitemOpen
  \bibfield  {author} {\bibinfo {author} {\bibfnamefont {A.}~\bibnamefont
  {Polman}}, \bibinfo {author} {\bibfnamefont {M.}~\bibnamefont {Knight}},
  \bibinfo {author} {\bibfnamefont {E.~C.}\ \bibnamefont {Garnett}}, \bibinfo
  {author} {\bibfnamefont {B.}~\bibnamefont {Ehrler}}, \ and\ \bibinfo {author}
  {\bibfnamefont {W.~C.}\ \bibnamefont {Sinke}},\ }\href {\doibase
  10.1126/science.aad4424} {\bibfield  {journal} {\bibinfo  {journal}
  {Science}\ }\textbf {\bibinfo {volume} {352}},\ \bibinfo {pages} {307}
  (\bibinfo {year} {2016})}\BibitemShut {NoStop}%
\bibitem [{\citenamefont {Fu}\ \emph {et~al.}(2017)\citenamefont {Fu},
  \citenamefont {Feldman}, \citenamefont {Margolis}, \citenamefont
  {Woodhouse},\ and\ \citenamefont {Ardani}}]{Fu2017}%
  \BibitemOpen
  \bibfield  {author} {\bibinfo {author} {\bibfnamefont {R.}~\bibnamefont
  {Fu}}, \bibinfo {author} {\bibfnamefont {D.}~\bibnamefont {Feldman}},
  \bibinfo {author} {\bibfnamefont {R.}~\bibnamefont {Margolis}}, \bibinfo
  {author} {\bibfnamefont {M.}~\bibnamefont {Woodhouse}}, \ and\ \bibinfo
  {author} {\bibfnamefont {K.}~\bibnamefont {Ardani}},\ }\href@noop {} {\emph
  {\bibinfo {title} {NREL}}},\ \bibinfo {type} {Tech. Rep.}\ \bibinfo {number}
  {September}\ (\bibinfo {year} {2017})\BibitemShut {NoStop}%
\bibitem [{\citenamefont {Green}(2016)}]{Green2016}%
  \BibitemOpen
  \bibfield  {author} {\bibinfo {author} {\bibfnamefont {M.~A.}\ \bibnamefont
  {Green}},\ }\href {\doibase 10.1038/nenergy.2015.15} {\bibfield  {journal}
  {\bibinfo  {journal} {Nature Energy}\ }\textbf {\bibinfo {volume} {1}},\
  \bibinfo {pages} {15015} (\bibinfo {year} {2016})}\BibitemShut {NoStop}%
\bibitem [{\citenamefont {Say}(1803)}]{say1803}%
  \BibitemOpen
  \bibfield  {author} {\bibinfo {author} {\bibfnamefont {J.-B.}\ \bibnamefont
  {Say}},\ }\href@noop {} {\emph {\bibinfo {title} {{Trait{\'{e}}
  d'{\'{e}}conomie politique}}}},\ edited by\ \bibinfo {editor} {\bibnamefont
  {Deterville}}\ (\bibinfo {year} {1803})\BibitemShut {NoStop}%
\bibitem [{\citenamefont {Hubbert}(1956)}]{Hubbert1956}%
  \BibitemOpen
  \bibfield  {author} {\bibinfo {author} {\bibfnamefont {M.~K.}\ \bibnamefont
  {Hubbert}},\ }\href@noop {} {\enquote {\bibinfo {title} {{Nuclear Energy and
  the Fossil Fuels}},}\ } (\bibinfo {year} {1956})\BibitemShut {NoStop}%
\bibitem [{\citenamefont {Sverdrup}\ \emph {et~al.}(2014)\citenamefont
  {Sverdrup}, \citenamefont {Koca},\ and\ \citenamefont
  {Ragnarsdottir}}]{Sverdrup2014}%
  \BibitemOpen
  \bibfield  {author} {\bibinfo {author} {\bibfnamefont {H.}~\bibnamefont
  {Sverdrup}}, \bibinfo {author} {\bibfnamefont {D.}~\bibnamefont {Koca}}, \
  and\ \bibinfo {author} {\bibfnamefont {K.~V.}\ \bibnamefont
  {Ragnarsdottir}},\ }\href {\doibase 10.1016/j.resconrec.2013.12.008}
  {\bibfield  {journal} {\bibinfo  {journal} {Resources, Conservation and
  Recycling}\ }\textbf {\bibinfo {volume} {83}},\ \bibinfo {pages} {121}
  (\bibinfo {year} {2014})}\BibitemShut {NoStop}%
\bibitem [{\citenamefont {Grandell}\ and\ \citenamefont
  {Thorenz}(2014)}]{Grandell2014}%
  \BibitemOpen
  \bibfield  {author} {\bibinfo {author} {\bibfnamefont {L.}~\bibnamefont
  {Grandell}}\ and\ \bibinfo {author} {\bibfnamefont {A.}~\bibnamefont
  {Thorenz}},\ }\href {\doibase 10.1016/j.renene.2014.03.032} {\bibfield
  {journal} {\bibinfo  {journal} {Renewable Energy}\ }\textbf {\bibinfo
  {volume} {69}},\ \bibinfo {pages} {157} (\bibinfo {year} {2014})}\BibitemShut
  {NoStop}%
\bibitem [{\citenamefont {Tao}\ \emph {et~al.}(2011)\citenamefont {Tao},
  \citenamefont {Jiang},\ and\ \citenamefont {Tao}}]{Tao2011}%
  \BibitemOpen
  \bibfield  {author} {\bibinfo {author} {\bibfnamefont {C.~S.}\ \bibnamefont
  {Tao}}, \bibinfo {author} {\bibfnamefont {J.}~\bibnamefont {Jiang}}, \ and\
  \bibinfo {author} {\bibfnamefont {M.}~\bibnamefont {Tao}},\ }\href {\doibase
  10.1016/j.solmat.2011.06.013} {\bibfield  {journal} {\bibinfo  {journal}
  {Solar Energy Materials and Solar Cells}\ }\textbf {\bibinfo {volume} {95}},\
  \bibinfo {pages} {3176} (\bibinfo {year} {2011})}\BibitemShut {NoStop}%
\bibitem [{\citenamefont {{Fraunhofer ISE}}(2015)}]{FraunhoferISE2015}%
  \BibitemOpen
  \bibfield  {author} {\bibinfo {author} {\bibnamefont {{Fraunhofer ISE}}},\
  }\href {\doibase 059/01-S-2015/EN} {\bibfield  {journal} {\bibinfo  {journal}
  {Agora Energiewende}\ ,\ \bibinfo {pages} {82}} (\bibinfo {year}
  {2015})}\BibitemShut {NoStop}%
\bibitem [{\citenamefont {Bleiwas}(2010)}]{Bleiwas2010}%
  \BibitemOpen
  \bibfield  {author} {\bibinfo {author} {\bibfnamefont {D.~I.}\ \bibnamefont
  {Bleiwas}},\ }\href {http://pubs.usgs.gov/circ/1365/Circ1365.pdf} {\bibfield
  {journal} {\bibinfo  {journal} {Usgs}\ }\textbf {\bibinfo {volume} {1365}},\
  \bibinfo {pages} {18} (\bibinfo {year} {2010})}\BibitemShut {NoStop}%
\bibitem [{\citenamefont {Graedel}(2011)}]{Graedel2011}%
  \BibitemOpen
  \bibfield  {author} {\bibinfo {author} {\bibfnamefont {T.}~\bibnamefont
  {Graedel}},\ }\href {\doibase 10.1146/annurev-matsci-062910-095759}
  {\bibfield  {journal} {\bibinfo  {journal} {Annual Review of Materials
  Research}\ }\textbf {\bibinfo {volume} {41}},\ \bibinfo {pages} {323}
  (\bibinfo {year} {2011})}\BibitemShut {NoStop}%
\bibitem [{\citenamefont {Grandell}\ \emph {et~al.}(2016)\citenamefont
  {Grandell}, \citenamefont {Lehtil}, \citenamefont {Kivinen}, \citenamefont
  {Koljonen}, \citenamefont {Kihlman},\ and\ \citenamefont
  {Lauri}}]{Grandell2016}%
  \BibitemOpen
  \bibfield  {author} {\bibinfo {author} {\bibfnamefont {L.}~\bibnamefont
  {Grandell}}, \bibinfo {author} {\bibfnamefont {A.}~\bibnamefont {Lehtil}},
  \bibinfo {author} {\bibfnamefont {M.}~\bibnamefont {Kivinen}}, \bibinfo
  {author} {\bibfnamefont {T.}~\bibnamefont {Koljonen}}, \bibinfo {author}
  {\bibfnamefont {S.}~\bibnamefont {Kihlman}}, \ and\ \bibinfo {author}
  {\bibfnamefont {L.~S.}\ \bibnamefont {Lauri}},\ }\href {\doibase
  10.1016/j.renene.2016.03.102} {\bibfield  {journal} {\bibinfo  {journal}
  {Renewable Energy}\ }\textbf {\bibinfo {volume} {95}},\ \bibinfo {pages} {53}
  (\bibinfo {year} {2016})}\BibitemShut {NoStop}%
\bibitem [{\citenamefont {Fustier}\ \emph {et~al.}(2017)\citenamefont
  {Fustier}, \citenamefont {Gray}, \citenamefont {Gundersen},\ and\
  \citenamefont {Hilboldt}}]{Fustier2017}%
  \BibitemOpen
  \bibfield  {author} {\bibinfo {author} {\bibfnamefont {B.~K.}\ \bibnamefont
  {Fustier}}, \bibinfo {author} {\bibfnamefont {G.}~\bibnamefont {Gray}},
  \bibinfo {author} {\bibfnamefont {C.}~\bibnamefont {Gundersen}}, \ and\
  \bibinfo {author} {\bibfnamefont {T.}~\bibnamefont {Hilboldt}},\ }\href@noop
  {} {\emph {\bibinfo {title} {HSBC Global Research}}},\ \bibinfo {type} {Tech.
  Rep.}\ \bibinfo {number} {September 2016}\ (\bibinfo {year}
  {2017})\BibitemShut {NoStop}%
\bibitem [{\citenamefont {Kerr}(2010)}]{Hoegh-guldberg2007}%
  \BibitemOpen
  \bibfield  {author} {\bibinfo {author} {\bibfnamefont {R.~A.}\ \bibnamefont
  {Kerr}},\ }\href@noop {} {\bibfield  {journal} {\bibinfo  {journal}
  {Science}\ }\textbf {\bibinfo {volume} {328}},\ \bibinfo {pages} {1624}
  (\bibinfo {year} {2010})}\BibitemShut {NoStop}%
\bibitem [{\citenamefont {Hughes}(2013)}]{Hughes2013}%
  \BibitemOpen
  \bibfield  {author} {\bibinfo {author} {\bibfnamefont {J.~D.}\ \bibnamefont
  {Hughes}},\ }\href {\doibase 10.1038/494307a} {\bibfield  {journal} {\bibinfo
   {journal} {Nature}\ }\textbf {\bibinfo {volume} {494}},\ \bibinfo {pages}
  {307} (\bibinfo {year} {2013})}\BibitemShut {NoStop}%
\bibitem [{\citenamefont {van~den Bergh}(2017)}]{VandenBergh2017a}%
  \BibitemOpen
  \bibfield  {author} {\bibinfo {author} {\bibfnamefont {J.~C. J.~M.}\
  \bibnamefont {van~den Bergh}},\ }\href {\doibase 10.1038/nclimate3113}
  {\bibfield  {journal} {\bibinfo  {journal} {Nature Climate Change}\ }\textbf
  {\bibinfo {volume} {7}},\ \bibinfo {pages} {107} (\bibinfo {year}
  {2017})}\BibitemShut {NoStop}%
\bibitem [{\citenamefont {Kallis}(2011)}]{Kallis2011}%
  \BibitemOpen
  \bibfield  {author} {\bibinfo {author} {\bibfnamefont {G.}~\bibnamefont
  {Kallis}},\ }\href {\doibase 10.1016/j.ecolecon.2010.12.007} {\bibfield
  {journal} {\bibinfo  {journal} {Ecological Economics}\ }\textbf {\bibinfo
  {volume} {70}},\ \bibinfo {pages} {873} (\bibinfo {year} {2011})}\BibitemShut
  {NoStop}%
\bibitem [{\citenamefont {Raftery}\ \emph {et~al.}(2017)\citenamefont
  {Raftery}, \citenamefont {Zimmer}, \citenamefont {Frierson}, \citenamefont
  {Startz},\ and\ \citenamefont {Liu}}]{Raftery2017}%
  \BibitemOpen
  \bibfield  {author} {\bibinfo {author} {\bibfnamefont {A.~E.}\ \bibnamefont
  {Raftery}}, \bibinfo {author} {\bibfnamefont {A.}~\bibnamefont {Zimmer}},
  \bibinfo {author} {\bibfnamefont {D.~M.~W.}\ \bibnamefont {Frierson}},
  \bibinfo {author} {\bibfnamefont {R.}~\bibnamefont {Startz}}, \ and\ \bibinfo
  {author} {\bibfnamefont {P.}~\bibnamefont {Liu}},\ }\href {\doibase
  10.1038/nclimate3352} {\bibfield  {journal} {\bibinfo  {journal} {Nature
  Climate Change}\ } (\bibinfo {year} {2017}),\
  10.1038/nclimate3352}\BibitemShut {NoStop}%
\bibitem [{\citenamefont {Jackson}(2009)}]{timjackson}%
  \BibitemOpen
  \bibfield  {author} {\bibinfo {author} {\bibfnamefont {T.}~\bibnamefont
  {Jackson}},\ }\href@noop {} {\emph {\bibinfo {title} {{Prosperity Without
  Growth: Economics for a Finite Planet}}}}\ (\bibinfo  {publisher}
  {Earthscan},\ \bibinfo {address} {London and New York},\ \bibinfo {year}
  {2009})\BibitemShut {NoStop}%
\bibitem [{\citenamefont {Tainter}\ and\ \citenamefont
  {Patzek}(2012)}]{Tainter2012}%
  \BibitemOpen
  \bibfield  {author} {\bibinfo {author} {\bibfnamefont {J.}~\bibnamefont
  {Tainter}}\ and\ \bibinfo {author} {\bibfnamefont {T.~W.}\ \bibnamefont
  {Patzek}},\ }\href@noop {} {\emph {\bibinfo {title} {{Drilling Down: The Gulf
  Oil Debacle and Our Energy Dilemma}}}}\ (\bibinfo  {publisher} {Springer},\
  \bibinfo {year} {2012})\BibitemShut {NoStop}%
\bibitem [{\citenamefont {Dussud}\ \emph {et~al.}(2016)\citenamefont {Dussud},
  \citenamefont {Guggemos},\ and\ \citenamefont {Riedinger}}]{Dussud2016}%
  \BibitemOpen
  \bibfield  {author} {\bibinfo {author} {\bibfnamefont {F.-S.}\ \bibnamefont
  {Dussud}}, \bibinfo {author} {\bibfnamefont {F.}~\bibnamefont {Guggemos}}, \
  and\ \bibinfo {author} {\bibfnamefont {N.}~\bibnamefont {Riedinger}},\
  }\href@noop {} {\emph {\bibinfo {title} {{Bilan {\'{e}}nerg{\'{e}}tique de la
  France pour 2015}}}},\ \bibinfo {type} {Tech. Rep.}\ (\bibinfo  {institution}
  {Service de L'observation et des statistiques (SOeS)},\ \bibinfo {year}
  {2016})\BibitemShut {NoStop}%
\end{thebibliography}%

\end{document}